# FUTURE OF WORK: ETHICS

David Pastor-Escuredo[1,2]
[1] UCL, UK
[2] LifeD Lab, Spain

**Abstract**

Work must be reshaped in the upcoming new era characterized by new challenges and the presence of new technologies and computational tools. Over-automation seems to be the driver of the digitalization process. Substitution is the paradigm leading Artificial Intelligence and robotics development against human cognition. Digital technology should be designed to enhance human skills and make more productive use of human cognition and capacities. Digital technology is characterized also by scalability because of its easy and inexpensive deployment. Thus, automation can lead to the absence of jobs and scalable negative impact in human development and the performance of business. A look at digitalization from the lens of Sustainable Development Goals can tell us how digitalization impact in different sectors and areas considering society as a complex interconnected system. Here, reflections on how AI and Data impact future of work and sustainable development are provided grounded on an ethical core that comprises human-level principles and also systemic principles.

**Future of work**

Technology and automation in the industry are not new (Noble, 2017; Orlikowski, 1992) but it is certainly gaining new traction with the advances of robotics, Data and Artificial Intelligence (Acemoglu & Restrepo, 2018; Madakam, Holmukhe, & Jaiswal, 2019; Romero, Bernus, Noran, Stahre, & Fast-Berglund, 2016). Each industrial revolution has implied a certain level of automation and shifts in jobs (Burnette, 2008), but the implication of new digital technologies and the so called Fourth Industrial Revolution (Schwab, 2017) are deeper and lead us to actually rethink the future of work and the role of humans (West, 2018; Willcocks & Lacity, 2016). Automation has resulted also in new types of jobs required for massive production and scalability. Some of these jobs have led to reduced creative skillsets. Digital automaton is certainly posing to replace operators with AI and robots at scale (Acemoglu & Restrepo, 2018; West, 2018; Willcocks & Lacity, 2016).

Digital Automation, implemented as the integration of technologies such as, but not only, Mobile Internet, Cloud computing, Internet of Things, Data, AI and Robotics, together with new processing schemas (i.e. edge computing (Satyanarayanan, 2017)) has become a labor disruption driver (OpendMind, 2020). Covid-19 surge has helped boost this disruption, defined as "a 'double-disruption' scenario for workers" (Forum, 2016, 2020). The scope of this disruption will not only include factories, but other physical spaces such as cities (Chen, Marvin, & While, 2020; Tiddi, Bastianelli, Daga, d'Aquin, & Motta, 2020) or farms (Asseng & Asche, 2019).

The spatial awareness and psychomotor skills of humans are still quite beyond the state of the art of robots (Torricelli et al., 2016), both in rigid and soft robotics (Manti, Cacucciolo, & Cianchetti, 2016). Nevertheless, the pace of new and better performing Machine Learning-based machines is increasing fast and we may be optimistic about robots taking over repetitive skills. The key issue will be the transition and reskilling of many workers a proper timing to avoid high rates of irreversible unemployment which will require specific policies.

There is an opportunity in digitalization to improve jobs quality and impact if digitalization is driven to achieve the Sustainable Development Goals (SDGs) such as gender equality, more decent work and sustainable industries and livelihoods, although risks exist (Griggs et al., 2013; Sachs, 2012; Vinuesa et al., 2020). Most part of these risks are associated with asymmetric transitions in several sectors putting many workers in vulnerable situations and increasing inequality. These risks are only magnified with the abrupt ongoing digitalized adaptation to COVID-19 (Forum, 2020).

Imagining how machines will reshape the future of work is not a new endeavor. In (Malone, 2004) it was proposed to envision future jobs from the lens of human-machines collaboration, interactions and deep organizational changes. Interactions and collaboration in systems imply thinking and understanding the collective level. *Collective Intelligence* emerges from groups when processes within groups result in more intelligence that the sum of individuals (Lévy & Bononno, 1997; Malone, Laubacher, & Dellarocas, 2009, 2010; Peach). The intersection of AI and *CI* is a promising new paradigm to drive human-machine interactions for a better society including work (Mulgan, 2018). Looking at machines from a behavioral perspective (Rahwan et al., 2019) and interacting with humans in groups, we can design new tools, systems and algorithms that we can name as *supermind*s (Malone, 2018). Here, we briefly discuss aspects of sustainable, inclusive, and ethical digitalization through AI and CI that would enable growth, scientific progress and also better well-being.

Digitalization is an opportunity for healthier and more sustainable cities (SDG-11) and can help redefining jobs, for instance, surveillance, maintenance and promotion of public spaces (Filipponi et al., 2010). Data and AI will be an excellent basis for designing and managing all aspects related to urban design in depth, energy supplies (SDG-7) and urban policy making. Transportation and distribution related jobs will be also greatly shaped by digital technologies, autonomous vehicles and tools to analyze mobility flows and real-time demand. Commerce, Restaurants, Hospitality, and Travel and Tourism are already undergoing a digital transformation where the physical and digital worlds will converge into new virtually-enhanced spaces that will transform the relationship with customers (SDG-9). Risk assessments and response to natural disasters (Pastor-Escuredo, Torres, Martínez-Torres, & Zufiria, 2020), epidemics (Martín-Calvo, Aleta, Pentland, Moreno, & Moro, 2020) and crisis based on Data and AI are becoming already a reality that will change resilience and response mechanisms and the action of public servants and emergency bodies. Human-centered cities will imply providing more services to citizens, searching for their engagement and also become more resilient and sustainable, for instance through decarbonization (SDG-7, SDG-13).

Rural areas can greatly improve their conditions through digitalization, attracting more population and business if work in the fields becomes more technological, digitally connected and resilient. Factories and manufacturers could increase their productivity with a relief of excessive work schedules for their workers in important sectors such as nutrition and fashion. That process of digitalization and robotization of physical spaces implies interactions between different SDGs such as sustainable communities (SDG-11), climate change (SDG-13), life in land (SDG-15), responsible consumption (SDG-12), hunger (SDG-2), poverty (SDG-1) and better work (SDG-8). Digitalization can really change the landscape of livelihoods in any region of the world producing deep changes in jobs demand and the need of skills, provided the economic sustainability and also the cultural acceptance. Furthermore, the environmental impact of digitalization will be an important matter as part of a sustainable and ethical future of work and livelihoods (SDG-15). More technological and scientific advances and a better understanding how machines and humans can interact in physical spaces are needed.

Although there are promising advances in robotics and Machine Learning algorithms specially based on Reinforcement Learning (Polydoros & Nalpantidis, 2017), it is remarkable that very dull tasks that humans are complex for AI. Until machines can perform fully automated tasks with safety and security requirements, human-in-the-loop solutions will be likely required. This suggests that the physical interaction of humans with things and the environment is a very unique characteristic of humans (and animals). Physical interaction and performance embody many mechanisms through our extremities, senses and nervous system. This fact suggests that future jobs should better exploit these unique skills and capabilities. This is not the trend as works still tend to be more centered on digital interfaces with machines that are available in fixed spaces. Enhancing physical experience and interactions in non-usual places is a way forwards for humans as species and to leverage their innate skills. Different digital interfaces that allow humans to work in different types of physical spaces can have a great impact in the use of cities and green spaces (SDG-11), physical and mental health and well-being (SDG-3) and life on land (SDG-15). In that sense, AI, Virtual Reality (Burdea & Coiffet, 2003; Tepper et al., 2017) and Internet of Things (Li, Da Xu, & Zhao, 2018; Ray, 2018) are the grounding technologies to develop "phygital" spaces where humans can carry out different types of socio-economic and commercial activity in deeper connection with the environment.

Digitalization should lead humans to interact better and more with the ecosystem. AI has shown promising applications for predictive analytics about environmental conditions and measuring the impact of different types of anomalies in livelihoods (Pastor-Escuredo et al., 2020; Zufiria et al., 2018). Data and AI can help communities to increase the efficiency and adaption of their livelihoods to climate change (SDG-13) and be more resilient to crisis and natural disasters (SDG-11) or situations of conflict (SDG-16). In that sense, migration and mobility related to labor markets which is a very significant social phenomenon in many countries around the globe could be better monitored. Labor-related migrations are a source of problems related to epidemics (SDG-3), conflicts and segregation (SDG-16), poverty and hunger (SDG-1 and SDG-2) and also vulnerability and inequalities (SDG-10), specially for women that in many cases have to carry on ensuring the safety and survival of households (SDG-5).

*Collective Intelligence* should ground digitalization to better interact with the ecosystem, approached from a systemic perspective. The specific tools that communities of farmers, manufacturers and suppliers require are yet to be designed helping to improve individual decision making (i.e. taking decision about hiring manpower or migrating to urban areas por job searching) but also better decisions that take into account the complexities of livelihoods (Scoones, 2009). This type of collective decision making may include deciding crops, sowing and harvesting strategies, community-level mobility and coping strategies or the creation of new partnerships and stakeholder engagement to leverage resources or financial stimulation.

A more interconnected management of industries and value chains requires better monitoring. Digital technologies can enable remote tasks minimizing risk and relieving from tiresome activities. Computer vision, Internet of Things and drones are technologies that can be leveraged for security and monitoring replacing humans in risky jobs. There are opportunities for better future of work in the sense that monitoring factories and work-spaces can help ensure that labor rights are enforced and promoted according to international regulations. Blockchain is a digital technology that enables digital traceability of financial and operational activities along the value chain that will facilitate measuring the impact of industries in social, financial, economic and environmental dimensions which will transform how organizations make decisions. Digital traceability has a great potential to ensure sustainable practices along the value chain (SDG-8 and SDG-9) and also facilitating trustful ecosystems between partners (SDG-17) helping improve evaluation within consortiums and commercial relationships. It is necessary to highlight that there is an intrinsic ethical risk of staff freedom and morale regarding all types of monitoring and surveillance. Monitoring systems have to be designed from ethical perspective and human-centered to ensure that they are respectful with privacy and promote that workers become more engaged and motivated.

Now looking at the individual and organization levels, cognitive skills will evolve in the digital era. The tasks that have traditionally performed through body-level physical interfaces are being adapted to virtual interfaces mainly driven by a deeper and more intensive vision and hand-driven manipulation. The capacity to ingest visual content at high speed is one of the main characteristics of digital citizens. These skills, as a complement of body-driven action can also be a source of imagination, integrative thinking and embodied visual analytics. Such skills are specially relevant for new systems of decision making that require the integration of high-dimensional spaces accounting for many variables and indicators. Holistic systems based on real-time data and IA algorithms and tools will likely become part of high-level management. The interaction and the navigation through high-dimensionality and complex data will be critical and will require new skills beyond data analytics. We can expect visual data analytics and sense making to be a more specialized task to interpret data and make decisions. Hybrid scientific and management teams are already becoming a reality in data-driven business and in emergencies teams managing complex processes such as epidemics.

However, complexity will increase not only because of internal information flows, but also because of their connectedness with the ecosystem. Decision making tools will be required across the skeleton of corporations to allow fast response to external stimuli, build up adaptation and resilience mechanisms and generate *Collective Intelligence.* Sensing and signaling mechanisms between parts of the organizations will be critical and those should be based on AI to avoid introducing a larger burden on managers and workers. In that sense, we can learn from biological systems that basic sensing mechanisms between cells generate harmonized functioning and development of tissues. The tissue analogy works properly to model how companies should work as an alternative to tight hierarchies. As in biological processes, tissues are multi-scale (Pastor-Escuredo & del Alamo, 2020) which means that connectedness and signaling can go beyond the organization level and help connecting organizations across sectors. It will be relevant for long-term sustainability and deep transformations to understand how organizations can communicate and collaborate at different levels using digital technologies (SDG-17). AI, Data and Blockchain will help measure better the transactions and value-interactions between organizations promoting transparency, collaboration and accountability.

Although human resources have already undergone a significant level of digitalization, profiling and matching algorithms of candidates and employees will keep growing, specially if private data is shared across social and professional networks. The risks in terms of privacy, discrimination and biases have been already warned, but AI, Data and digital tools are still a great opportunity to build human-based tissues within organizations to increase creativity, variety and motivation. In the last years, we have witnessed the creation of networks of experts and stewards to promote collaboration and data collaboratives (Verhulst, Zahuranec, Young, & Winowatan, 2020). These are good models of how new capacities and roles will be created within organizations to harness the potential of digital-based collaboration and new connections in value chains.

In the same way we start monitoring devices and physical systems through with the Internet of Things and other digital technologies, there is a potential risk in the instrumentalization of people and machines for the sake of efficiency and performance. Human empowering CI is a proper framework to evaluate how digitalization should be developed to create and manage teams beyond the existing tools for collaboration and creation of intellectual or industrial assets (Malone, 2018; Malone et al., 2010). Behavioral science, psychology and augmented CI are elements that have to be integrated to properly design the internal digitalization of teams and the creation of superminds. *Co-creation* is a critical element of team building and collaboration and can be greatly improved by integrating the ideas of teams with tools to search and check evidence and also nowcasting the needs and trends of the society using Big Data and AI. Thus, digitalization is an opportunity for democratized and shared ideas and projects that are also well connected with the real-world and are evidence-based. Besides, immersive virtual environments will change how remote work is done, allowing distant collaboration and work to be much more effective. Measuring impact of projects, teams and partnerships will be also critical at all scales. Impact assessment tools are yet to be developed beyond the current KPIs frameworks. Holistic frameworks that account for the dynamics and synergies of groups and the SDGs will help promoting more decent work and more sustainable organizations.

As discussed, it can be argued that AI and Data biases will be a great problem than a solution (Kusner & Loftus, 2020). However, existing corporations are not free of biases. Cognitive biases are frequent as well as biased generated by personal interests and power relationships. Another role of AI that can shape the future of work is as a mediator and as a catalyzer to change power structures within corporations. Transparent and accountable algorithms are now driven many of customer-oriented decisions of many platforms that were natively digital or have gone digital in the recent years. We can expect algorithms to make decisions affecting internal organization and managing management. Decentralized Autonomous Organizations built on Blockchain and smart contracts are an example of new paradigms of algorithmic organization. This paradigm encompasses a clear risk that machines will eventually be the organizers and managers ruling humans for the sake of efficiency and efficacy. Hybrid human-machine systems designed from the lens of CI can potentially be the solution to this wicked problem.

Eventually, the real ethical question and scientific challenge is if machines can help humans be more fair in their decisions, increase their awareness and knowledge to avoid biases, have a more holistic and comprehensive vision of needs for a sustainable society and self-regulate from self-interest (Laffont & Martimort, 2009; Pastor-Escuredo & Vinuesa, 2020).

Avoiding the risks for ethical and sustainable future world demands proactive actions from policies and financial instruments to specific research and collaboration-based ecosystems. AI design has to be integrated with ethical and SDGs-driven frameworks to envision human-machine systems that empower workers and ease the transition to a more digitalized society where workers will need new skillsets. However, in most cases, trends of digitalization are leading to scenarios of vulnerability, rushed up by the COVID-19 pandemia. It is urgent to design and implement new systems and superminds that help build resilience and drive changes within organizations, sectors and livelihoods to open new spaces and opportunities for all type of workers.

*Acknowledgements. Authors are indebted to the MIT/MISTI-Spain Seed Funds "Empowering Collective Intelligence With Artificial Intelligence To Enhance And Scale Sustainable Development (Sustainable Cities)" that partially support the preparation of this paper.*